\documentstyle [floats,preprint,epsf,aps,eqsecnum]{revtex}

\global\arraycolsep=2pt		

\makeatletter
\def\footnotesize{\@setsize\footnotesize{9.5pt}\xpt\@xpt
\abovedisplayskip 10pt plus2pt minus 5pt
\belowdisplayskip \abovedisplayskip
\abovedisplayshortskip \z@ plus 3pt
\belowdisplayshortskip 6pt plus 2pt minus 2pt
\def\@listi{\topsep 6pt plus 2pt minus 2pt
\parsep 3pt plus 2pt minus 1pt \itemsep \parsep}}
\makeatother

\newcommand{\fnref}[1]{\mbox{$^{\ref{#1}}$}}
\newcommand{\eqnref}[1]{Eq.~(\ref{#1})}

\def\bfbeta{\hbox{\setbox0=\hbox{$\beta$}%
   \kern-.025em\copy0\kern-\wd0
   \kern.05em\copy0\kern-\wd0
   \kern-.025em\raise.0433em\box0}}

\begin {document}

\preprint {UW/PT-96-08}

\title{Compton Scattering on Black Body Photons}

\author{Lowell S. Brown and Ronald S. Steinke}

\address{ Department of Physics, \\
    University of Washington,
    Seattle, Washington 98195 \\
    }%
\maketitle

\begin {abstract}
    {%
    We examine Compton scattering of electrons on black body photons in
    the case where the electrons are highly relativistic, but the center
    of mass energy is small in comparison with the electron mass. 
    We derive the partial lifetime of electrons in the LEP accelerator due to
    this form of scattering in the vacuum beam pipe and compare it with
    previous results.
    }%
\end {abstract}


%

\newpage

\section{Introduction}

Vacuum beam pipes of modern particle accelerators closely approach the
ideal limit of a pipe completely devoid of gas molecules. However,
even an ideal vacuum beam pipe in a laboratory at room temperature is
filled with photons having an energy distribution given by Planck's
law. Some time ago, Telnov\fnref{Telnov} noted that the scattering of
electrons on these black body photons could be a significant mechanism
for the depletion of the beam.  
This scattering of the electrons
in the Large Electron Positron collider at CERN (LEP) on the black
body radiation has been detected$^{2\hbox{--}4}$. There is a long
history of theoretical investigations on the scattering of high-energy
electrons on black body photons, centering around the role this plays
as an energy loss mechanism for cosmic rays, which is
summarized by Blumenthal and Gould\fnref{Blumenthal+Gould}. More 
recently, Domenico\fnref{Domenico} and Burkhardt\fnref{Burkhardt} have
considered this effect for the LEP experiments and the consequent
limit on the beam lifetime by using numerical Monte Carlo methods. In
view of the intrinsic interest of the problem of high-energy electron
scattering on black body photons, we believe that it is worthwhile to
present here a simplified calculation of the effect. We compute the
total cross section analytically. The cross section as a function of
the energy loss --- which is the important quantity for the beam
lifetime --- is also done analytically except for a final
straightforward numerical integration. Our calculations use
relativistic invariant methods, and are thus of some pedagogical
interest.

Analytic computations can be performed because the problem involves
two small dimensionless parameters. On the one hand, the electron of
mass $m$ has a very large laboratory energy $E$ and it is
ultrarelativistic, as characterized by the parameter $ m^2 / E^2
$. (We use natural units in which the velocity of light $c=1$,
Planck's constant $ \hbar = 1 $, and Boltzmann's constant $ k =1 $, so
that temperature is measured in energy units.)  At LEP, $ m^2 / E^2
 \approx 10^{-10} $. We shall neglect terms of order $ m^2 / E^2 $. On the
other hand, the temperature $ T $ of the black body radiation is very
small in comparison with the electron mass $ m $. Thus, although the
electron is ultrarelativistic, the energy in the center of mass of the
electron-photon system is still small in comparison with the electron
mass. The head-on collision of a photon of energy $ T $ with an
relativistic electron of energy $ E $ produces, with the neglect of
the electron mass, the squared center-of-mass energy $ 4ET $. We
shall use the dimensionless parameter (which gives an average
value)
\begin{equation} 
s = 2ET / m^2 \,.  
\end{equation} 
At LEP, $ s \approx 10^{-2} $. Thus it
is a good approximation to use the non-relativistic limit in the 
center of mass, with the relativistic Compton cross section replaced 
by its constant, non-relativistic Thomson limit. To assess this 
approximation, we shall also compute the first corrections in 
$ s $. 

In the next section we use simple relativistic techniques to compute
the total cross section for the scattering of an ultrarelativistic
electron on the Planck distribution of black body photons. The third
section describes the more detailed calculation needed for the cross
section in which the electron loses an energy greater than $ \Delta E
$. If the energy loss $ \Delta E $ in an electron-photon collision is
too large, the electron's motion  
falls outside of the
acceptance parameters of the machine. 
At LEP this happens when $ \Delta E / E $ is
greater than about 1\%. As we shall see, this means that even if the
beam were in a perfect vacuum, it would decay with a half life of
about two days. The vacuum in the LEP accelerator is so good that the
beam scattering of the black body photons is, in fact, the primary
mechanism for beam loss when the machine is run with a single
beam. Scattering on the residual gas in the beam pipe gives a
considerably longer half life of about six days.\fnref{BandK} When
the machine is run in the usual mode with two beams for $e^+ \, e^-$
collision experiments, beam-beam collisions reduce the beam half life
to about 14 hours.\fnref{BandK}

\section{Total Scattering Rate}
In the general case of an electron scattering off some photon distribution,
the scattering rate $\bar \Gamma$ in the electron's rest frame may be
computed using the formula
\begin{equation}
\bar \Gamma = \int \frac{(d^3\bar k)}{(2\pi)^3} \bar f({\bf \bar k})
\sigma({\bf \bar k}) \,,
\label{restscatter}
\end{equation}
where $f({\bf \bar k})$ is the photon phase-space density [with the
normalization defined such that $\bar j^0$ in \eqnref{jmu} is the photon
number density] as a function of the photon
momentum and $\sigma({\bf \bar k})$ is the scattering cross section, which
is similarly a function of the photon momentum. Here all quantities are
evaluated in the electron's rest frame as indicated by the over
bar. This scattering rate may be viewed as a time derivative
\begin{equation}
\bar \Gamma = { d n \over d\tau } \,,
\end{equation}
where $\tau$ is the time in the electron's rest frame. Since numbers
are Lorentz invariant and $\tau$ may be defined to be the invariant
proper time of the electron, the rate $dn/d\tau$ is, in fact, a
Lorentz invariant. Thus, if the integral on the right-hand side of the
rate formula (\ref{restscatter}) is written in a Lorentz invariant
manner, we can immediately evaluate the rate in the laboratory
frame. In the lab frame, the electron moves with four momenta
\begin{equation}
p^\mu = m { dz^\mu \over d\tau} \,,
\end{equation}
where $ z^\mu(\tau)$ is the world line of this particle, its
space-time position as a function of proper time, and $m$ is the
electron mass.  This gives the familiar time-dilation formula
\begin{equation}
{dt \over d\tau} = { d z^0 \over d\tau} = { p^0 \over m} = {E \over m}
\,,
\end{equation} 
where $E$ is the electron's total energy. Thus the scattering rate
$\Gamma$ in the lab frame may be easily evaluated using
\begin{equation}
\Gamma = { dn \over dt} = { m \over E } { dn \over d\tau} \,.
\end{equation}

In the non-relativistic limit,
$\sigma$ may be replaced with the Thomson cross section, $\sigma_T =
8\pi r_0^2 / 3$, where $r_0 = e^2/4\pi m$ is the classical electron radius.
Since this is
independent of ${\bf k}$, the scattering rate may be rewritten as
\begin{equation}
\bar \Gamma_0 = \sigma_T \, \bar j^0 \,,
\end{equation}
where
\begin{equation}
\bar j^{\mu} = \int \frac{(d^3 \bar k)}{(2\pi)^3} \frac{\bar k^{\mu}}{\bar k^0}
\bar f({\bf \bar k})
\label{jmu}
\end{equation}
is the photon number flux four vector. Since $ \bar p^\mu / m = (1\,,
{\bf 0} )$ in the electron's rest frame, we may write this leading
approximation, denoted with a $0$ subscript,  as
\begin{equation}
\left( \frac{dn}{d\tau} \right)_0 =
 -\sigma_T \, \bar j^{\mu}\, \frac{\bar p_{\mu}}{m} \,,
\label{invar}
\end{equation}
with the minus sign arising from our Lorentz metric convention in
which the metric has the signature $(-,+,+,+)$. The result
(\ref{invar}) is now in an invariant form which holds in any
frame. With a thermal photon distribution in the lab frame, 
\begin{equation}
f(k) = \frac{2}{e^{\omega/T}-1} \,,
\label{number}
\end{equation}
where $\omega = k^0$ is the photon energy, the photon number
distribution is isotropic, and so only the number density component $j^0$ is
nonvanishing. Thus, in the lab frame,
\begin{equation}
\Gamma_0 =  \frac{m}{E} \,
\sigma_T \, j^0 \, \frac{p^0}{m} = \sigma_T \, j^0 \,.
\label{gammawithj}
\end{equation}
The lab photon number density obtained from integrating (\ref{jmu})
with the distribution (\ref{number}) is the familiar result 
\begin{equation}
j^0 = \frac{2\zeta(3)}{\pi^2} T^3 \,,
\end{equation}
in which $\zeta(3) = 1.202\dots$ is the Riemann zeta function. 
Thus, $\Gamma_0$ is given by
\begin{equation}
\Gamma_0 = \frac{2\zeta(3)}{\pi^2}T^3\sigma_T \,.
\label{gamma0}
\end{equation}

The first order relativistic correction to this result is obtained
with the use of the corrected cross section 
\begin{equation}
\sigma = \sigma_T \left( 1 + \frac{2pk}{m^2} \right) \,.
\end{equation}
Note that the product $kp = k^\mu p_\mu$ of the two four-momenta is
negative with our metric. 
Because $\sigma$ is no longer 
independent of $k$, the corresponding form of \eqnref{gammawithj} is
\begin{eqnarray}
\Gamma_1 &=& \frac{m}{E} \sigma_T
\int \frac{(d^3k)}{(2\pi)^3} \, \frac{-kp}{k^0m} \, f(k) \left( 1
+ \frac{2pk}{m^2} \right)
\nonumber \\
&= & -\frac{1}{E} \, \sigma_T 
\left( j^{\mu}p_{\mu} + \frac{2}{m^2} T^{\mu\nu}
p_{\mu}p_{\nu} \right) \,,
\end{eqnarray}
where
\begin{equation}
T^{\mu\nu} = \int \frac{(d^3k)}{(2\pi)^3} \frac{k^{\mu}k^{\nu}}{k^0} f( k)
\label{tmunu}
\end{equation}
is the stress-energy tensor of the photons. Due to the isotropy of the
thermal photons in the lab frame, $T^{\mu\nu}$ has no off diagonal components,
and it is also traceless because the photon is massless, $k^\mu k_\mu
= 0$. Therefore, in the lab frame,
\begin{equation}
T^{\mu\nu}p_{\mu}p_{\nu} = \left( E^2 + \frac{1}{3}|{\bf p}^2| \right) T^{00}
= \frac{4}{3}E^2 \left( 1 - \frac{m^2}{4E^2} \right) T^{00} \,.
\end{equation}
The $m^2/E^2$ term is very small, and it may be neglected.
Integrating over the photon distribution in \eqnref{tmunu} gives the
well-known black body energy density
\begin{equation}
T^{00} = \frac{6\zeta(4)}{\pi^2} T^4 \,,
\end{equation}
where $ \zeta(4) = \pi^4 / 90 = 1.082\dots$.  This yields the
corrected  scattering rate
\begin{equation}
\Gamma_1 = \frac{2\zeta(3)}{\pi^2} T^3 \sigma_T \left[ 1
- 4s\frac{\zeta(4)}{\zeta(3)} \right]\,.
\label{firstcorrate}
\end{equation}

For the temperature in the LEP beam pipe we take
$T=291\,\hbox{K}=0.0251\,\hbox{eV}$, which is about room
temperature. This gives the leading rate
$\Gamma_0 = 9.98\times10^{-6}\,\hbox{s$^{-1}$}$ corresponding to the
mean life $\tau_0 = 1/\Gamma_0 = 28$ hr. A typical LEP beam energy
$E=46.1\,\hbox{GeV}$ is just above half the $Z^0$ mass --- within the
width, but on the high side of resonance curve. Together with the
previous value of the temperature, this gives $ s = 0.00886$, and the
first-order corrected rate $\Gamma_1 =
9.66\times10^{-6}\,\hbox{s$^{-1}$}$, which is about 3\% smaller than
the leading rate.  This gives a mean life $\tau_1 = 1/\Gamma_1
= 29\,\hbox{hr}$.

\section{Rate With Energy Loss}

The calculation of the scattering rate in which the electron loses
an energy greater than $ \Delta E $ is facilitated by going back to
the basic formula\fnref{Brown} that expresses the rate in terms of Lorentz
invariant phase space integrals, an energy-momentum conserving $
\delta $ function, and the square of the scattering amplitude $ |T|^2
$.  The total electron scattering rate as observed in the lab frame reads
\begin{equation}
\Gamma  = \frac{1}{2E} \int \frac{(d^3k)}{(2\pi)^3}
\frac{1}{2\omega} f(k) \int \frac{(d^3k')}{(2\pi)^3} \frac{1}{2\omega'}
\int \frac{(d^3p')}{(2\pi)^3} \frac{1}{2E'} (2\pi)^4 \delta^{(4)}(k'
+ p' - k - p) |T|^2 \,,
\label{tot}
\end{equation}
where $p$ and $p'$ are the initial and final electron four momenta,
$k$ and $k'$ the initial and final photon four momenta, with $ E = p^0
\,,\, E' = p^{\prime \, 0} \,,\, \omega = k^0 \,,\, 
\omega' = k^{\prime \, 0} $ the time components of these four vectors. 
Except for the initial factor of $1/2E$ which is the lab energy of the
initial electron and which converts the invariant
proper time into the lab time, the right-hand side of this
expression is a Lorentz invariant. The problem proves to be greatly
simplified if the integrals are evaluated in the rest
frame of the electron, 
because Compton scattering of a photon on an electron at rest has a
very simple nonrelativistic limit. This complicates the initial photon
distribution, but, if we introduce a four vector $\beta^{\mu}$, whose time
component in the lab frame is one over the temperature of the photon
distribution and whose spatial components are zero in the lab frame,
the distribution in an arbitrary frame still has the simple form
\begin{equation}
f(k) = \frac{2}{e^{-\beta^{\mu}k_{\mu}}-1} \,.
\label{dist}
\end{equation}
From the definition of $\beta^{\mu}$,  $-\beta^{\mu}
\beta_{\mu} = 1/T^2$ and $-\beta^{\mu}p_{\mu} = E/T$, because multiplication
by $\beta^{\mu}$ selects out the time component in the lab frame. In the
electron rest frame, $\beta^{\mu}$ therefore takes on the value
\begin{equation}
\beta^{\mu} = \left( \frac{E}{Tm} , -\frac{{\bf p}}{Tm} \right) \,,
\label{betadef}
\end{equation}
where $E$ and ${\bf p}$ are taken in the lab frame. We have not yet taken into
account the lower bound on the electron energy loss in the lab frame. Because
multiplication by $\beta^{\mu}$ selects the time component in the lab frame,
this limit may be instituted by the inclusion of an ``energy loss"
step function in the integrand of Eq.~(\ref{tot}), 
\begin{equation}
\theta \left( -\beta^{\mu}(p_{\mu} - p'_{\mu}) - \frac{\Delta E}{T} \right) \,,
\end{equation}
where the $1/T$ in the second term has been included to compensate for
the factor of $1/T$ which the first term has picked up by being multiplied
by $\beta^{\mu}$.
Using the identity
\begin{equation}
\int \frac{(d^3p')}{(2\pi)^3} \frac{1}{2E'} = \int \frac{(d^4p')}{(2\pi)^4}
\theta(E' - m) 2\pi \delta(p'^2 + m^2) \,,
\end{equation}
the final electron four momentum $p'$ may be integrated over to leave
\begin{eqnarray}
\Gamma(\Delta E) &=& \frac{1}{2E} \int \frac{(d^3k)}{(2\pi)^3}
\frac{1}{2\omega} f(k) \int \frac{(d^3k')}{(2\pi)^3} \frac{1}{2\omega'}
\theta(\omega - \omega') 2\pi \delta(-2m\omega + 2m\omega' - 2kk')
\nonumber \\
& & \qquad \theta \left( -\beta k' - \left( \frac{\Delta E}{T}  - \beta k
\right) \right) |T|^2 \,.
\label{good}
\end{eqnarray}

We do the ${\bf k'}$ integral in spherical coordinates and take the
$z$ axis to be parallel to ${\bf k}$, with $\theta$ the angle between
these two vectors. The angle $\theta$ is the photon scattering angle
in the electron rest frame, and
\begin{equation}
-kk' = \omega \omega' (1-\cos\theta) \,.
\end{equation}
The $\delta$ function can now be solved for $\omega'$ to yield
\begin{equation}
\delta(-2m\omega + 2m\omega' - 2kk') = \frac{1}{2(m + \omega(1 - \cos\theta))}
\delta \left( \omega' - \frac{m\omega}{m + \omega(1 - \cos\theta)} \right) \,,
\label{energy}
\end{equation}
which requires that $\omega' < \omega$ and thus makes the $\theta (
\omega - \omega')$ step function redundant. The 
scattered photon energy $\omega'$ given by the $\delta$ function is,
of course, just the Compton energy. To deal with the ``energy loss"
step function, we note that the Lorentz transformation from the lab
frame to the initial electron rest frame turns the lab frame isotropic
black body photon distribution into a very narrow pencil in the
electron rest frame in which we are now working. Thus the initial
photon distribution is sharply peaked about the average value
\begin{equation}
\overline{(k^{\mu}/\omega)} = \beta^{\mu}/\beta^0 \,.
\end{equation}
Hence we can approximate
\begin{equation}
-\beta k' \simeq \frac{\beta^0}{\omega} (-kk') = \beta^0 \omega'
(1 - \cos\theta) \,.
\end{equation}
To verify this and assess the order of accuracy, we define the average
more precisely by  
\begin{equation}
\langle X \rangle =
\frac{\int \frac{(d^3k)}{(2\pi)^3} \frac{1}{2\omega}
f(-k\beta) \, X }
{\int \frac{(d^3k)}{(2\pi)^3} \frac{1}{2\omega} f(-k\beta) } \,.
\end{equation}
Then, by virtue of the relativistic invariance of this definition,
\begin{equation}
\langle k^\mu k^\nu \rangle = \left( \beta^\mu \beta^\nu - {1 \over
4} \beta^2 g^{\mu\nu} \right) A(\beta^2) \,,
\end{equation}
since $\beta^\lambda$ is the only four vector available and $ k^\mu
k_\mu = 0$. Remembering that $\omega = k^0$, this presents the squared
fluctuation about the average as
\begin{equation}
\frac{\left\langle \left( k^{\mu} - \beta^{\mu} \frac{\omega}{\beta^0} \right)
\left( k^{\nu} - \beta^{\nu} \frac{\omega}{\beta^0} \right) \right\rangle}
{\left\langle\omega^2\right\rangle} = B \left( g^{\mu\nu}
- \frac{\beta^{\nu}}{\beta^0} g^{\mu 0} - \frac{\beta^{\mu}}{\beta^0} g^{\nu 0}
+ \frac{\beta^{\mu}\beta^{\nu}}{(\beta^0)^2} \right)  \,,
\end{equation}
where
\begin{equation}
B = { - \beta^2 \over 4(\beta^0)^2 + \beta^2}
 \simeq \frac{m^2}{4E^2} \,.
\end{equation}
Thus the deviations away from our approximation may be neglected
because they involve the very small quantity $ m^2 / E^2$. 
 Using this
approximation for $-k'\beta$ simplifies the ``energy loss" step
function to
\begin{equation}
\theta \left( \omega' - \frac{\Delta E}{\beta^0 T (1 - \cos\theta)} \right)
= \theta \left( \omega' - \frac{\Delta E}{E} \frac{m}{1-\cos\theta} \right) \,,
\end{equation} 
where the $-\beta k$ in the original step function has been neglected
because it is much less than $\Delta E/T$. Inserting the value of 
$\omega'$ given by the energy-conserving $\delta$ function
(\ref{energy}) into the step function gives
\begin{equation}
\theta \left( \frac{m\omega}{m + \omega(1-\cos\theta)} - \frac{\Delta E}{E}
\frac{m}{1-\cos\theta} \right) = \theta \left( \omega - \frac{\Delta E}{E'}
\frac{m}{1-\cos\theta} \right) \,,
\end{equation}
where on the right hand side we have solved for $\omega$ and defined
\begin{equation}
E' = E - \Delta E \,,
\end{equation}
which is the maximum final electron energy in the lab frame.

We perform the ${\bf k}$ integral in spherical coordinates, with the
polar angle $\chi$  taken to be the angle
between ${\bf k}$ and \bfbeta, so that 
\begin{equation}
-k\beta = \omega( \beta^0 - |{\bfbeta}|\cos\chi ) \,.
\end{equation}
We rewrite the angular integral for ${\bf k}$ in terms of an integration
over $ k\beta $ by noting the limits
\begin{equation}
-k\beta < \omega(\beta^0 +|{\bfbeta}|) \simeq 2\omega \beta^0  = {
2 \omega E \over m T } \,,
\end{equation}
and
\begin{equation}
-k\beta > \omega(\beta^0 - |{\bfbeta}|) = \frac{\omega \beta^2}{\beta^0
+ |{\bfbeta}|} \simeq \frac{\omega m}{2TE} \,.
\end{equation}
Thus, with the neglect of order $ m^2 / E^2$ corrections, and
remembering that $|\bfbeta| \simeq E / mT $, 
\begin{equation}
\int_{-1}^1 d\cos\chi = \frac{mT}{\omega E}
\int_{\omega m/2TE}^{2\omega E/mT} d(-k\beta) \,.
\end{equation}
We shall do the $-k\beta$ integral last, due to its dependence on the
initial photon distribution. In order to interchange the order of the
$\omega$ and $ -k\beta$ integrations, we note that
the lower limit on $-k\beta$,
\begin{equation}
-k\beta > \frac{\omega m}{2TE} \,,
\end{equation}
gives the upper bound on $\omega$,
\begin{equation}
\omega < \frac{2TE(-k\beta)}{m} = s (-k\beta) m \,.
\end{equation}
The upper  limit on $-k\beta$,
\begin{equation}
-k\beta < \frac{2\omega E}{mT} \,,
\end{equation}
gives the lower bound on $\omega$,
\begin{equation}
\omega > \frac{(-k\beta)mT}{2E} = s (-k\beta) m { m^2 \over 4 E^2} \,.
\end{equation}
In view of the extreme smallness of $ m^2 / E^2 $, we may replace this
lower limit by $ \omega = 0$. Hence, switching the order of integration gives
\begin{equation}
\int_0^{\infty} d\omega \int_{\omega m/2TE}^{2\omega E/mT} d(-k\beta)
= \int_0^{\infty} d(-k\beta) \int_0^{s (-k\beta) m} d\omega \,.
\end{equation}
We perform this reversal of integrals, do the two trivial azimuthal
integrals, and do the $\omega'$ integral using the $\delta$ function
to obtain
\begin{eqnarray}
\Gamma(\Delta E) &=& \frac{m^2T}{16E^2(2\pi)^3} \int_0^{\infty}
d(-k\beta) f(-k\beta) \int_{-1}^1 d\cos\theta
\int_0^{s (-k\beta) m } d\omega
\nonumber \\
& & \frac{\omega}{[m + \omega(1-\cos\theta)]^2} \, |T|^2 \,
\theta\left( \omega - \frac{\Delta E}{E'} \frac{m}{1-\cos\theta} \right) \,.
\end{eqnarray}

To work out the integrals which appear here, it is convenient to first
introduce the appropriate, dimensionless variables,
\begin{equation}
x = -k\beta \,, \qquad z = 1 - \cos\theta \,, \qquad \nu = { \omega
\over s (-k\beta) m } \,,
\end{equation}
and define
\begin{equation}
u = { \Delta E \over 2 s E' } \,.
\label{par}
\end{equation}
With this new notation, we have
\begin{equation}
\Gamma(\Delta E) = \frac{T^3}{4m^2(2\pi)^3} \int_0^{\infty}
dx \, x^2 f(x) \int_0^2 dz \int_0^1 d\nu
 \frac{\nu}{(1 + s \nu xz )^2} \, |T|^2 \,
\theta\left( \nu - { 2 u \over xz}  \right) \,.
\end{equation}
The final step function provides the lower limit  $\nu = 2u / xz$.
This lower limit must not exceed the upper limit $ \nu =
1 $. Hence we must have condition $ z > 2u/x$ on the $z$
integration. But again, this must not exceed the upper limit $ z =
2$. Thus $ x > u$, and imposing all these limits gives
\begin{equation}
\Gamma(\Delta E) = \frac{T^3}{4m^2(2\pi)^3} \int_u^{\infty}
dx \, x^2 f(x) \int_{2u/x}^2 dz \int_{2u/xz}^1 d\nu \,
 \frac{\nu}{(1 + s \nu xz )^2} \, |T|^2 \,.
\label{great}
\end{equation}

This will be evaluated in the nonrelativistic limit, keeping first order
corrections in $s$. The exact squared amplitude differs from its
nonrelativistic limit
\begin{equation}
|T|^2 = 2e^4(1+\cos^2\theta) = 2e^4(2-2z+z^2)
\end{equation}
by corrections of order $ \omega \omega' / m^2$. These corrections
involve $s^2$ and are thus negligible. To first order in $s$,
\begin{equation}
\frac{1}{(1 + s \nu xz)^2} \simeq 1 - 2 s \nu xz  \,.
\end{equation}
The $z$ and $\nu$ integrations are now straightforward. We express
the result as
\begin{equation}
\Gamma(\Delta E) = \Gamma_0 \left[ F_0(u) - 4s\frac{\zeta(4)}{\zeta(3)}
F_1(u) \right] \,,
\end{equation}
where $\Gamma_0$ is the approximate total scattering rate from \eqnref{gamma0}.
The straightforward integrations give
\begin{equation}
F_0(u) = \frac{1}{2\zeta(3)} \int_u^{\infty} \frac{dx}{e^x-1} \left( x^2 - 3ux
+ 3u^2\ln\left(\frac{x}{u}\right) + \frac{2u^3}{x} \right) \,,
\label{zero}
\end{equation}
and\fnref{check}
\begin{equation}
F_1(u) = \frac{1}{6\zeta(4)} \int_u^{\infty} \frac{dx}{e^x-1} \left( x^3
- \frac{9}{2} u^2x - u^3 + 6u^3\ln\left(\frac{x}{u}\right) + \frac{9u^4}{2x}
\right) \,.
\label{one}
\end{equation}
It can be seen that in the $\Delta E \to 0$ limit, $F_0(0) = F_1(0) = 1$,
so $\Gamma(\Delta E)$ reduces to the result (\ref{firstcorrate}) for
$\Gamma_1$. 

At this stage, one must resort to numerical calculations to evaluate
the integrals. However, analytic calculations of the energy weighted
moments of the distribution can still be performed. The simplest of
these is the average energy loss observed in the lab frame, 
$\langle E-E' \rangle$. 
Because $\Gamma(\Delta E)$ describes the rate due to all scattering
events where $E - E' > \Delta E$, this average 
value may be computed by
\begin{equation}
\langle E-E' \rangle = - \int_0^{\infty} d\Delta E \Delta E \frac{d}{d\Delta E}
\left( \frac{\Gamma(\Delta E)}{\Gamma_1} \right) \,,
\end{equation}
where $\Gamma_1$ is the total scattering rate including the first
correction in $s$ given by Eq.~(\ref{firstcorrate}). Changing variables
to $u$ and integrating by parts gives
\begin{equation}
\langle E-E' \rangle  = \int_0^{\infty} du 
\frac{\Gamma(\Delta E)}{\Gamma_1}
\frac{d\Delta E}{du} \,,
\end{equation}
with, in view of Eq.~(\ref{par}), 
\begin{equation}
\frac{d\Delta E}{du} = \frac{2sE}{(1+2su)^2} \,.
\end{equation}
Inserting the expressions for $\Gamma(\Delta E)$ and $\Gamma_1$
into the integral and expanding in powers of $s$ gives
\begin{equation}
\langle E-E' \rangle = 2sE \int_0^{\infty} du \left\{ F_0(u)
- 4s \left[ u F_0(u) - \frac{\zeta(4)}{\zeta(3)} F_0(u)
+ \frac{\zeta(4)}{\zeta(3)} F_1(u) \right] + O(s^2) \right\} \,.
\label{mom}
\end{equation}
Inserting the expressions for the $F$'s from \eqnref{zero} and
\eqnref{one} and interchanging the order of the $x$ and $u$ integrals,
the integrals may be evaluated analytically, and we find that
\begin{eqnarray}
\langle E-E' \rangle &=& 2sE \frac{\zeta(4)}{\zeta(3)} \left\{ 
1 + s \left[ 4
\frac{\zeta(4)}{\zeta(3)} - \frac{63}{5} \frac{\zeta(5)}{\zeta(4)}
\right] + O(s^2) \right\} 
\nonumber\\
&&
\nonumber\\
&=& 1.80 \, s E \Big[ 1 - 8.5 \, s \Big]
\,.
\label{loss}
\end{eqnarray}

To check that no mistakes have been made in our calculation of
$F_0(u)$ and $F_1(u)$ given in 
\eqnref{zero} and \eqnref{one}, we have independently evaluated the
average energy loss $\langle E-E' \rangle$ starting from
Eq.~(\ref{good}) and only making the small $s$ approximation towards
the end of the calculation. We find the same result with this
different method. 

\begin{figure}
  \epsfxsize=5.0in
   \epsfysize=4.0in
  \leavevmode {\hfill \epsfbox{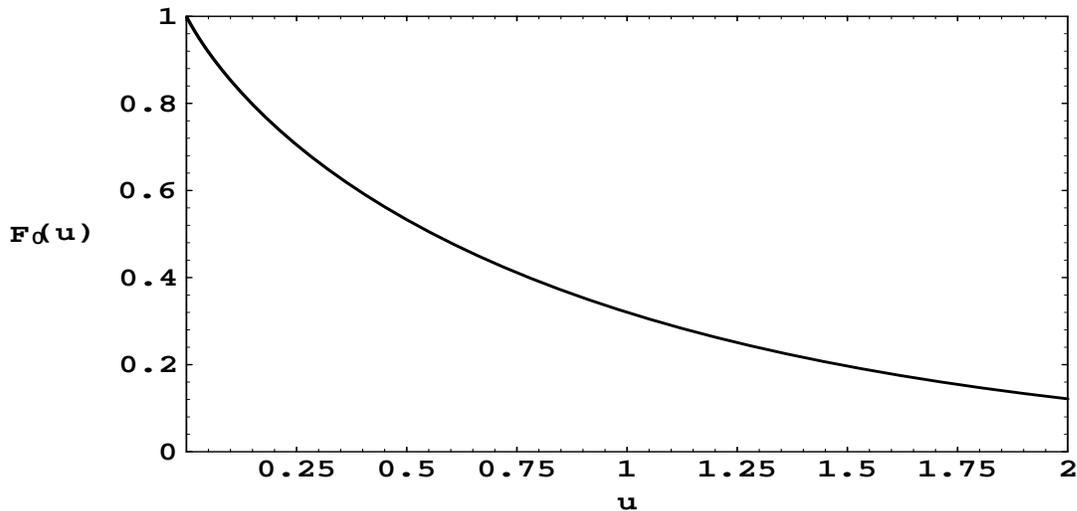} \hfill}                 
             
  \caption{The dimensionless $F_0(u)$ defined in Eq.~(3.35) as a
function of the dimensionless variable $u$ defined in Eq.~(3.29). }
  \label{rateplot}
\end{figure}

\begin{figure}
   \epsfxsize=5.0in
   \epsfysize=4.0in
   \leavevmode {\hfill \epsfbox{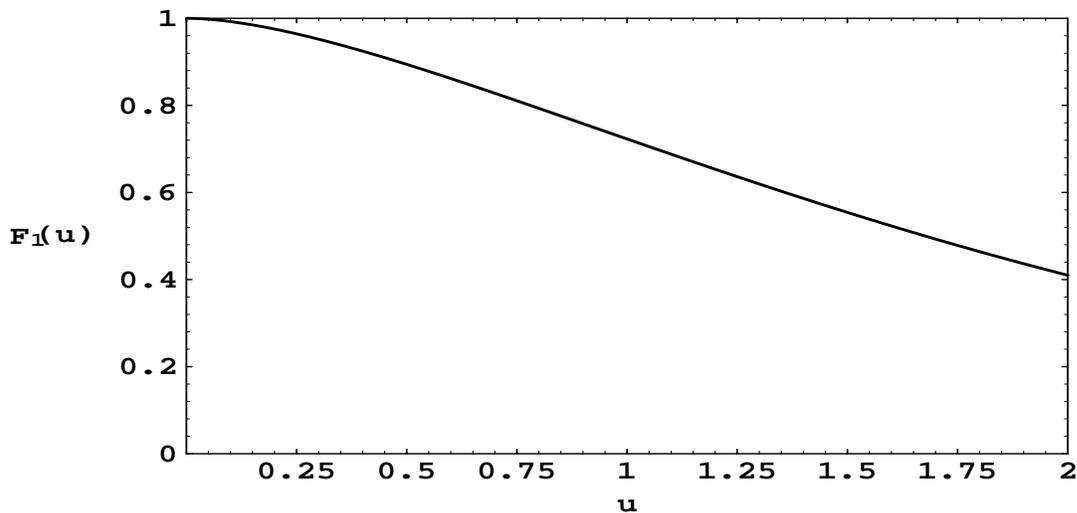} \hfill}

   \caption{The dimensionless $F_1(u)$ defined in Eq.~(3.36) as a
function of $u$ .}
   \label{firstcor}
\end{figure}

The integrals in the definitions (\ref{zero}) and (\ref{one}) of 
the functions $F_0(u)$ and $F_1(u)$ have been calculated
numerically, and the results are displayed in Fig. \ref{rateplot} and
Fig. \ref{firstcor}. As a check on this numerical result, we have
used it to evaluate the integrals in Eq.~(\ref{mom}) numerically, and
the results agree with the analytic expression (\ref{loss}) to within
$0.2$ percent.

We may compare our calculations with the those of
Domenico\fnref{Domenico} who employed a Monte Carlo method. He
 used the values $E = 46.1$ GeV and $ T = 291$ K (which we have
previously employed) that give $ s = 0.0089$. He also took
$\Delta E = 0.012 E$ which places $u = 0.69$. 
Numerical integration gives $F_0(0.69) = 0.44$ and
$F_1(0.69) = 0.83$, and from these values we calculate a mean beam
lifetime of 64 hours to zeroth order in the nonrelativistic
limit, and of 68 hours 
when the first order relativistic corrections are
included. This is to be compared with Domenico's value of 90 hours for the
same input parameters. We do not understand the reason for this discrepancy.

We may also compare our results with those of
H. Burkhardt\fnref{Burkhardt} who also used a Monte Carlo method.
Burkhardt uses parameters that are slightly different than those used
by Domenico, namely $E = 45.6$ GeV and $T = 295$ K. These parameters
yield the same $s = 0.0089$. However, since the overall rate scales as
$T^3$, which is 4\% larger with Burkhardt's temperature, our value of
the beam lifetime for $\Delta E = 0.012 E$ with his parameters is
reduced from 68 to 65 hours. Burkhardt finds 83 hours. (To compare
with Domenico, we note that modifying his result of 90 hours by the
4\% change in $T^3$ produces 86 hours.) Burkhardt and
Kleiss\fnref{BandK} also state that the average fractional energy loss
$ \langle E - E' \rangle / E$ is 1.1\% for this value of $s$, but our
result (\ref{loss}) gives the larger value 1.5\% corresponding to our
shorter beam lifetime. Again, we we can only state that we do
not understand the reason for these discrepancies.

\begin{center}

ACKNOWLEDGMENTS

\end{center}

We should like to thank J. Rothberg for making us aware of the
electron scattering on the photons in the LEP beam pipe. 
We would also like to thank
H. Burkhardt for informing us of relevant literature and C. Woll 
for checking much of our calculations. This work was
supported, in part, by the U.S. Department of Energy under
Grant. No. DE-FG03-96ER40956.

\newpage

\begin{center}
REFERENCES
\end{center}

\begin{enumerate}

\item{\label{Telnov} V. I. Telnov, ``Scattering of electrons on thermal
radiation photons in electron-positron storage rings,''
Nucl. Instr. and Meth. {\bf A260}, 304--308 (1987).}

\item{\label{Dehning} B. Dehning, A. C. Melissinos, F. Perrone,
C. Rizzo and G. von Holtey, ``Scattering of high
energy electrons off thermal photons,''  Phys. Lett. {\bf B
249}, 145--148 (1990).}

\item{\label{Bini1} C. Bini, G. De Zorzi, G. Diambrini-Palazzi,
G. Di Cosimo, A. Di Domenico, P. Gauzzi and D. Zanello, ``Scattering
of thermal photons by a 46 GeV positron beam at LEP,'' Phys. Lett.
{\bf B 262}, 135--138 (1991).}

\item{\label{Bini2} C. Bini, G. De Zorzi, G. Diambrini-Palazzi,
G. Di Cosimo, A. Di Domenico, P. Gauzzi and D. Zanello, ``Fast
measurement of luminosity at LEP by detecting the single
bremsstrahlung photons,'' Nucl. Instr. and Meth. {\bf A306},
467--473 (1991).}

\item{\label{Blumenthal+Gould} G. R. Blumenthal and R. J. Gould, 
``Bremsstrahlung, synchrotron radiation, and Compton scattering
of high-energy electrons traversing dilute gases,'' Rev. Mod. Phys
{\bf 42}, 237--270 (1970).}

\item{\label{Domenico} A. D. Domenico, ``Inverse Compton scattering
of thermal radiation at LEP and LEP-200,'' Part. Accel. {\bf 39},
137--146 (1992).}

\item{\label{Burkhardt} H. Burkhardt, ``Monte Carlo Simulation of Beam
Particles and Thermal Photons'', CERN/SL Note 93-73 (OP), 1993
(Internal Note, unpublished). }

\item{\label{BandK} See, for example, 
H. Burkhardt and R. Kleiss, ``Beam Lifetimes in LEP'', Proc. 4th
Eur. Part. Acc. Conf. EPAC, London 1994, Eds. V. Suller and
Ch. Petit-Jean-Genaz, Vol II, pp. 1353-1355.} 

\item{\label{Brown} See, for example, L. S. Brown, {\it Quantum Field
Theory} (Cambridge Univ. Press, 1992), Section 3.4.}

\item{\label{check} The result for $F_0(u)$ given in Eq.~(\ref{zero})
can be obtained from an integration of Eq.~(2.42) of Blumenthal and
Gould\fnref{Blumenthal+Gould}, but the result for $F_1(u)$ in
Eq.~(\ref{one}) appears to be new.}

\end{enumerate}

\end{document}